# On the Compressed Measurements over Finite Fields: Sparse or Dense Sampling

Jin-Taek Seong and Heung-No Lee[*]
School of Information and Communications,
Gwangju Institute of Science and Technology (GIST), Gwang-ju, South Korea
Email: {jtseong, heungno[*]}@gist.ac.kr

**Abstract**

We consider compressed sampling over finite fields and investigate the number of compressed measurements needed for successful $L_0$ recovery. Our results are obtained while the sparseness of the sensing matrices as well as the size of the finite fields are varied. One of interesting conclusions includes that unless the signal is "ultra" sparse, the sensing matrices do not have to be dense.

I. INTRODUCTION

In recent years, the Compressed Sensing (CS) theory has emerged, and it allows us to provide a new signal acquisition paradigm in which compression and sampling of signals can be done simultaneously, introduced in the signal processing and information theory literature, such as Candes and Tao [1] and Donoho [2]. One of the main issues in the CS problems has been to quantify how many measurements are needed for perfect recovery of the unknown signals. The most surprising and interesting discovery is that perfect recovery is possible with a number of measurements much smaller than the ambient dimension of the unknown signal, as long as the signal being sampled is sufficiently sparse.

In general, the problem of CS has been considered mainly in the field of real and complex systems. One of the key points in CS problems is to minimize the number of measurements while unknown signals are perfectly recovered. In this paper, we aim to find recovery conditions for CS problems over finite fields. There are some applications that this problem can be useful, including, *i*) the problem of collecting data samples from a group of correlated sources [5] and [6], *ii*) group testing [7], *iii*) the problem of sensor failure detection [8], *iv*) minimization of file servers to contact in order to complete a download in a file sharing network [9]. For instance, in [5], Bassi et al. addressed the problem of the collecting spatially correlated measurements in a wireless sensor network. All sensors quantize their measurements, and map them to *q*-level symbols. The sink receives coded packets which are linearly combined from Galois Field arithmetic operations with source packets. The reconstruction of source packets can be done via solving an underdetermined linear equation over finite fields. The core of these problems is the one that we consider in this paper. There are a couple of related works. Draper and Malekpour [3] reported on the error exponents for recovery of sparse signals using uniform random sensing matrices over finite fields. Tan et al. have extended the works of [3] to the problem of rank minimization [4], and showed that the minimum rank decoder achieves the information theoretic lower bound as long as the fraction of nonzero entries of the sensing matrices scales as $\Omega\left(\frac{\log N}{N}\right)$ where $N$ denotes the size of signals.

In this paper, we aim to investigate the core question of CS problems again, but for the CS systems over finite fields where the sparse signals, the sensing matrix, and the measurements are all made of the elements from a finite field of a certain size. We use the ideal $L_0$ recovery routine with a goal of providing benchmark for practical recovery routines. We first consider *dense* sensing matrices and then *sparse* ones as well, and investigate the impact of sparseness in sensing matrices on uniqueness in unknown sparse signal recovery. One interesting result is that sparse sensing matrices are as good as dense ones unless the signal of interest is "ultra" sparse.

## II. Compressed Sampling over Finite Fields

We describe the following system model in the finite field of the size $q$ as $\mathbb{F}_q$: Let $\mathbf{x} \in \mathbb{F}_q^N$ be a signal vector of length $N$ with sparsity $k_1$, which indicates the number of nonzero entries from 0 to $K$ in $\mathbf{x}$, $k_1 \in \{0,1,\ldots,K\}$, and let $\mathbf{A} \in \mathbb{F}_q^{M \times N}$ be an $M \times N$ sensing matrix with $N > M$. The measured signal $\mathbf{y}$ is

given as

$$\mathbf{y} = \mathbf{A}\mathbf{x}. \qquad (1)$$

Let $\mathcal{L}$ denote the set, as $\mathcal{L} := \bigcup_{k_1=0}^{K} \mathcal{L}_{k_1}$ where $\mathcal{L}_{k_1}$ denotes the set of signals of length $N$ with sparsity $k_1$. And the size of the set $\mathcal{L}$ is given by $|\mathcal{L}| = \sum_{k_1=0}^{K} \binom{N}{k_1}(q-1)^{k_1}$, where $|\cdot|$ denotes the cardinality of the set. The sparse signal is randomly and uniformly selected from the set $\mathcal{L}$. We assume that the elements of the sensing matrix $\mathbf{A}$ are independent and identically distributed (i.i.d.), so that

$$\Pr\{A_{ij} = \alpha\} = \begin{cases} 1-\gamma & \alpha = 0, \\ \gamma/(q-1) & \alpha \neq 0, \end{cases} \qquad (2)$$

where $\gamma$ denotes the sparse factor of the sensing matrix, $A_{ij}$ denotes the element of the $i$-th row and the $j$-th column of the sensing matrix where $i = 1, 2, \ldots, M$ and $j = 1, 2, \ldots, N$, and $\alpha$ denotes the dummy variable such as $\alpha \in \mathbb{F}_q$. For uniform random sensing matrices, the sparse factor is given as $1 - q^{-1}$. With respect to the sparseness of the sensing matrix, we investigate the recovery performance of a CS framework for the given parameters, i.e., $N$, $K$, $M$, and $\gamma$ over finite fields.

## III. Uniform Random Sensing Matrices

In this section, we derive the necessary and sufficient conditions for uniform random sensing matrices on the unique recovery of sparse signals. We assume that the decoder in our scheme finds a sparsest feasible solution $\hat{\mathbf{x}}$ using the $L_0$ minimization rule,

$$(L_0) \quad \min \|\bar{\mathbf{x}}\|_0 \quad \text{subject to} \quad \mathbf{A}\bar{\mathbf{x}} = \mathbf{y}, \qquad (3)$$

where $\bar{\mathbf{x}} \in \mathcal{L}$ is a feasible solution. Note that the sparity of the $L_0$ recovery, denoted as $k_2 := \|\hat{\mathbf{x}}\|_0$, is less than or equal to the sparsity $k_1$ of the unknown signal $\mathbf{x}$. An error is said to have occurred when a feasible but not exactly the unknown vector $\bar{\mathbf{x}}$ is decided by the $L_0$ recovery, i.e., $\mathbf{y} = \mathbf{A}\bar{\mathbf{x}}$ but $\mathbf{x} \neq \bar{\mathbf{x}}$. The error event of this decoder is $\mathcal{E}_0 := \{\hat{\mathbf{x}} \neq \mathbf{x}\}$. We found the analysis based on this error event is rather difficult and thus work on the following error event,

$$\mathcal{E} := \{(\mathbf{A}, \mathbf{x} \in \mathcal{L}, \bar{\mathbf{x}} \in \mathcal{L}): \hat{\mathbf{x}} \neq \mathbf{x}, \mathbf{y} = \mathbf{A}\hat{\mathbf{x}}\}. \tag{4}$$

Note that $\mathcal{E}_0 \subseteq \mathcal{E}$ and thus $\Pr\{\mathcal{E}_0\} \leq \Pr\{\mathcal{E}\}$. Then, the probability of error is upper bounded by

$$\begin{aligned}\Pr\{\mathcal{E}_0\} &\leq \Pr\{\mathcal{E}\} \\ &= \frac{1}{|\mathcal{L}|} \sum_{\mathbf{x} \in \mathcal{L}} \sum_{\substack{\bar{\mathbf{x}} \in \mathcal{L} \\ \hat{\mathbf{x}} \neq \mathbf{x}}} \Pr\{\mathbf{A}\mathbf{x} = \mathbf{A}\bar{\mathbf{x}}\}.\end{aligned} \tag{5}$$

It is noteworthy that (5) is almost intractable to evaluate since $|\mathcal{L}|$ is typically very large. This brute-force approach can be avoided with what will be described subsequently here. Namely, we enumerate all elements of the set $\mathcal{L}$ by providing indices $l_1$ and $l_2$ from 0 to $|\mathcal{L}|-1$ such as $\{\mathbf{x}_0, \mathbf{x}_1, \ldots, \mathbf{x}_{|\mathcal{L}|-1}\}$ and $\{\bar{\mathbf{x}}_0, \bar{\mathbf{x}}_1, \ldots, \bar{\mathbf{x}}_{|\mathcal{L}|-1}\}$. Let $\mathbf{d}_{l_1 l_2}$ denote the difference vector between $\mathbf{x}_{l_1}$ and $\bar{\mathbf{x}}_{l_2}$. By dividing all the vectors represented by $\mathbf{d}_{l_1 l_2}$ into smaller sets which have the same Hamming weights of $h$, (5) can be rewritten as

$$\Pr\{\mathcal{E}\} = \frac{1}{|\mathcal{L}|} \sum_{h=1, l_1 \neq l_2}^{2K} N_h \Pr\{\mathbf{A}\mathbf{d}_h = 0\}, \tag{6}$$

where $N_h$ denotes the number of difference vectors with $\|\mathbf{d}_{l_1 l_2}\|_0 = h$, $\mathbf{d}_h$ denotes a vector with $\|\mathbf{d}_{l_1 l_2}\|_0 = h$ for $h = 1, 2, \ldots, 2K$. Our approach is utilizing the fact the probabilities are the same within each group: namely, if $\|\mathbf{d}_{l_1 l_2}\|_0 = \|\mathbf{d}_{l'_1 l'_2}\|_0 = h$ for any $l_1 \neq l'_1 \in \mathcal{L}$ and $l_2 \neq l'_2 \in \mathcal{L}$, then $\Pr\{\mathbf{A}\mathbf{d}_{l_1 l_2} = 0\} = \Pr\{\mathbf{A}\mathbf{d}_{l'_1 l'_2} = 0\}$. That is, for any $\beta \in \mathbb{F}_q \setminus \{0\}$ and $\alpha \in \mathbb{F}_q$, $\Pr\{A_{ij}\beta = \alpha\} = \Pr\{A_{ij} = \alpha\}$. It is to be noted that since the elements of the sensing matrix are i.i.d., then, $\Pr\{\mathbf{A}\mathbf{d}_h = 0\} = \prod_{i=1}^{M} \Pr\{A_i\mathbf{d}_h = 0\}$ where $A_i$ denotes the $i$-th row of $\mathbf{A}$. We compute the probability $\Pr\{A_i\mathbf{d}_h = 0\}$ as follows,

$$\Pr\{A_i \mathbf{d}_h = 0\} \overset{(a)}{=} \Pr\left\{\sum_{j=1}^{h} A_{ij} = 0\right\} = q^{-1}, \qquad (7)$$

where the equality (*a*) is from the fact that the vector $\mathbf{d}_h$ has exactly $h$ nonzero elements and the elements of uniform random sensing matrices defined in (2) are i.i.d.. Now, (6) can be rewritten as

$$\begin{aligned}
\Pr\{\mathcal{E}\} &= \frac{1}{|\mathcal{L}|} \sum_{h=1, l_1 \neq l_2}^{2K} N_h q^{-M} \\
&\overset{(a)}{=} (|\mathcal{L}| - 1) q^{-M} \\
&\leq K 2^{NH_b(K/N)} 2^{K \log_2(q-1)} 2^{-M \log_2(q)},
\end{aligned} \qquad (8)$$

where $H_b(\cdot)$ denotes the binary entropy, and the equality (*a*) originates from the fact that $\sum_{h=1, l_1 \neq l_2}^{2K} N_h = (|\mathcal{L}| - 1) |\mathcal{L}|$. Consequently, from the condition that the exponent of (8) remains negative so that the probability of error goes to 0 as $N \to \infty$, we can derive the following upper bound on *M*,

$$M \geq \frac{NH_b(K/N) + K \log_2(q-1)}{\log_2 q}. \qquad (9)$$

**Proposition 1** (*Sufficient Condition*). Let $\gamma = 1 - q^{-1}$ in (2). If (9) is satisfied, then $\Pr\{\mathcal{E}\} \to 0$ as $N \to \infty$.

Next, we derive the necessary condition for the unique recovery of sparse signal. For this, we aim to consider the Markov chain relation, since decision $\hat{\mathbf{x}}$ is made given $\mathbf{A}$ and $\mathbf{y}$, i.e., $\mathbf{x} \to (\mathbf{A}, \mathbf{y}) \to \hat{\mathbf{x}}$, a standard approach in information theory. Then, by the Fano's inequality [13], the probability of error $\Pr\{\mathcal{E}_0\}$ is lower bounded as follows,

$$\Pr\{\mathcal{E}_0\} \geq \frac{H(\mathbf{x}|\mathbf{y}, \mathbf{A}) - 1}{\log_q |\mathcal{L}|}, \qquad (10)$$

where $H(\cdot)$ denotes the entropy. According to the definition of conditional entropy, $H(\mathbf{x}|\mathbf{y}, \mathbf{A}) = H(\mathbf{x}) - I(\mathbf{x}; \mathbf{y}, \mathbf{A})$ where $I(\cdot)$ denotes the mutual information. Assuming that $\mathbf{A}$ is independent of $\mathbf{x}$, we have $I(\mathbf{x}; \mathbf{y}, \mathbf{A}) = I(\mathbf{x}; \mathbf{y} | \mathbf{A})$. We use the following

$I(\mathbf{x};\mathbf{y}|\mathbf{A}) = I(\mathbf{y};\mathbf{x}|\mathbf{A}) = H(\mathbf{y}|\mathbf{A}) - H(\mathbf{y}|\mathbf{x},\mathbf{A})$. Since $\mathbf{y}$ is a function of $\mathbf{A}$ and $\mathbf{x}$, $H(\mathbf{y}|\mathbf{x},\mathbf{A}) = 0$. Subsequently, (10) can be rewritten as

$$\Pr\{\mathcal{E}_0\} \geq \frac{H(\mathbf{x}) - H(\mathbf{y}|\mathbf{A}) - 1}{\log_q |\mathcal{L}|}. \tag{11}$$

Since $H(\mathbf{y}|\mathbf{A}) \leq H(\mathbf{y}) \leq M$, we obtain the lower bound,

$$\Pr\{\mathcal{E}_0\} \geq \frac{H(\mathbf{x}) - M - 1}{\log_q |\mathcal{L}|}. \tag{12}$$

**Proposition 2** (*Necessary Condition*). Let $\gamma = 1 - q^{-1}$ in (2). If

$$M < \log_q \left[ (q-1)^K \binom{N}{K} \right] - 1, \tag{13}$$

then $\Pr\{\mathcal{E}_0\} > 0$ as $N \to \infty$.

From both Propositions 1 and 2, we observe the limit conditions for recovery of sparse signals over finite fields. In fact, for large values of $N$, the two conditions (9) and (13) converge.

## IV. Sparse Random Sensing Matrices

The CS theory shows that sparse signals can be perfectly recovered if the sensing matrix satisfies the so-called Restricted Isometry Property (RIP). Most widely proposed matrices satisfying RIP are Gaussian and Fourier matrices. They are matrices with full density. The computational load in signal sampling and signal recovery with dense sampling matrices, however, grows exponentially fast with the ambient dimension $N$ of signals. This motivates one to consider the use of sparse sensing matrices, a research topic which in fact has received considerable interest recently. Sparse sensing matrices are shown as good as the dense ones, satisfying an RIP [10], while with much reduced signal processing complexity both in signal sampling and in signal recovery [10]-[12].

In this section, we aim to investigate the recovery performance using *sparse* sensing matrices. For this, we extend the analysis given in Section III for the system in which the sensing matrix is also *sparse* in addition to the signal being *sparse*.

The sparsity condition on sensing matrix is incorporated into (7) in the following way, and the rest of the steps leading to (6) are the same, i.e.,

$$\Pr\{A_i \mathbf{d}_h = 0\} \overset{(a)}{=} q^{-1} + (1-q^{-1})\left(1 - \frac{\gamma}{1-q^{-1}}\right)^h, \tag{14}$$

where the equality (*a*) is obtained by taking the *h*-fold circular convolution of the probability distribution (2).

Next, we compute $N_h$. For this, we use a combinatorial approach which is to enumerate all difference vectors into mutually exclusive groups each with the same Hamming weight. We assume that $\mathbf{x}_{l_1}$ has the first $k_1$ elements with nonzeros and the rest of the $N - k_1$ elements are zeros. We solve this problem in three steps. First, we consider the set $\mathcal{L}_{k_1}$ of exact signals with sparsity $k_1$. Second, we enumerate all candidate signals $\bar{\mathbf{x}}$ with sparisty $k_2$ corresponding to the same Hamming weights of the difference vector. A candidate signal has $t$ nonzeros in the second set of indices, i.e., $\{k_1+1, k_1+2, \cdots, N\}$, and the rest $k_2 - t$ nonzeros in the first set of indices, i.e., $\{1, 2, \cdots, k_1\}$. It is to be noted that $t \in \{0, 1, \cdots, k_2\}$ and $2t \leq h \leq k_2 + t$. That is, for a given $t$, $h$ is at least $2t$ since if there are $t$ differences in the first index set, then there should be the same number of differences in the second index set; $h$ is $(k_2 + t)$ at maximum when $(k_2 - t)$ nonzero elements of $\bar{\mathbf{x}}$ in the first set is not equal to 1, viz. $(k_2 - t)$ the number of differences in addition to the $2t$ minimum differences. Third, similar to the set $\mathcal{L}_{k_1}$ we have just considered, all the other sets of exact signals should be considered in the same manner described in the first two steps here. Finally, the number of difference vectors with $\|\mathbf{d}_l\|_0 = h$ can be expressed as follows,

$$N_h = \sum_{k_1=0}^{K}\sum_{k_2=0}^{k_1}\sum_{t=0}^{k_2}\binom{k_1}{k_2-t}\binom{k_2-t}{h-2t-(k_1-k_2)}(q-2)^{h-2t-(k_1-k_2)}\binom{N-k_1}{t}(q-1)^t. \tag{15}$$

Using (14) and (15), we can obtain the upper bound on the probability of error for *sparse* matrices. From the upper bound, we investigate the impact of sensing matrix sparseness of for a unique recovery of a sparse signal.

## V. Discussions

Fig. 1 shows the plot of the compression ratio (=$M/N$) versus the sparsity ratio (=$K/N$) from 0 to 0.5 with the signal length $N = 1000$. We consider the size of finite fields as follows: $q = 2$, 4, 16, and 256. This plot is drawn from the expression given in (6), such that for a fixed $K$, the smallest integer $M$ is found for the probability of error (6) less than $10^{-2}$. One interesting feature of this figure is that when the size of the finite field increases, the compression ratio (or the under-sampling ratio) becomes small. That is, the number of measurements required for unique recovery of unknown sparse signals dramatically decreases. For example, for recovery of signals with sparsity ratio of 0.2, the compression ratio is 0.72, 0.51, 0.38, and 0.29 for $q = 2$, 4, 16, and 256, respectively. In addition, the numbers of measurements with respect to different sparse factors are obtained. Two classes of sparse factors are considered: one is for dense matrices, i.e., the sparse factor is $1 - q^{-1}$, and the other is $\frac{C \cdot \log N}{N}$ where $C$ is a constant that can vary the sparseness of sensing matrices. Let us consider the case for $C = 10$. In the example of $q = 4$, the sparse factors are as follows: for uniform random sensing matrices, $\gamma = 0.75$; however, for sparse random sensing matrices with $C = 10$, $\gamma \approx 0.069$. The sparse factor in the latter case is very small in comparison with that of the first. This difference indicates that the sensing matrix generated from the sparse factor in the second case is sparser by a factor of about 10 when compared to that of the uniform sensing matrix. Although a large difference exists between the sparse factors of the sensing matrices, their recovery performances are nearly identical except for the *ultra*-sparse region, say $\frac{K}{N} < 0.1$. These behaviors are observed over all the finite fields shown in fig. 1. The results imply that the sparse sensing matrices are comparable to the dense matrices in terms of recovery performance. There are related works [10] and [11] in the real-valued CS systems in which the research aim is to investigate if sparse sensing matrices work as good as dense ones do. They have shown that sparse ones satisfy an RIP condition as well as the dense ones do.

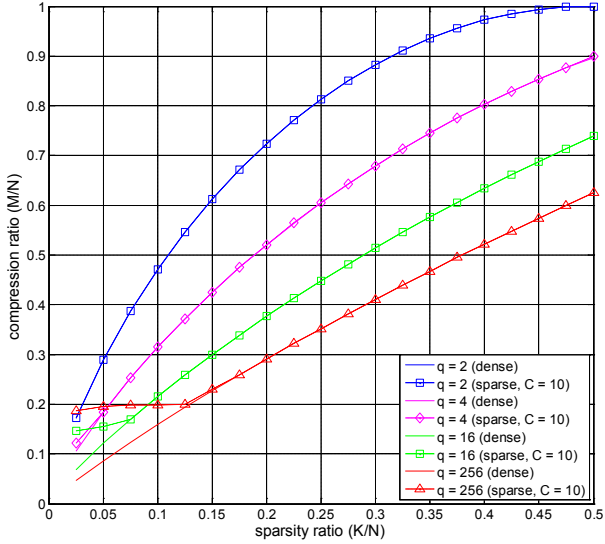
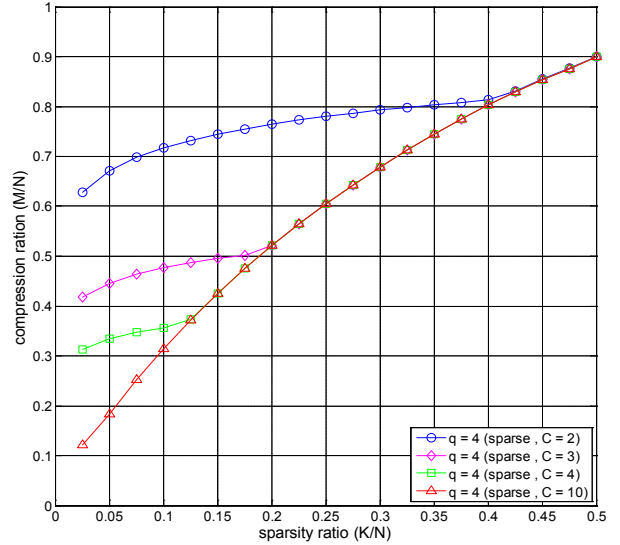

Fig. 1. Comparisons of the probability of error expression (6) (evaluated at $10^{-2}$) with dense and sparse sensing matrices for $N = 1000$: solid lines (dense, $\gamma = 1 - 1/q$), dotted lines (sparse, $\gamma \approx C \times 0.0069$).

Fig. 2 Comparisons of results of the probability of error (6) less than $10^{-2}$ with various sparse factors for $q = 4$ and $N = 1000$.

From fig. 2, it can be observed that in the ultra-sparse region, a higher value of sparse factor $\gamma$ is required to recover the signals. It is easy to see that if the signal and the sensing matrix are both sparse, the chance of having zero valued compressed sample goes very high; a consequence to this effect is the increase in the required number of measurements so as to compensate for the number of missed sensing opportunities. The purpose of fig. 2 is to show that as the sparse factor of sensing matrix increases, the curve of the error probability approaches to that of the dense sensing matrices even in the ultra-sparse region.

## VI. CONCLUSIONS

In this work, we considered a framework of CS over finite fields. We investigated the recovery performance for two classes of sensing matrices. For uniform random sensing matrices, we obtained the necessary and sufficient conditions for a unique recovery of a sparse signal with high probability. In addition, we showed that the recovery performance using sparse sensing matrices can be made as good as that of the dense ones over various finite sizes. We found that for recovery of ultra-sparse signals not too sparse sensing matrices are required. From our main results, we provided answer to the key question: how many number of measurements are needed in the compressed sensing system over finite fields.